\documentclass[%
 reprint,
 amsmath,amssymb,
 aps,
]{revtex4-1}

\usepackage{graphicx,float}
\usepackage{dcolumn}
\usepackage{bm}
\usepackage{subfigure}
\usepackage{braket}
\usepackage{xcolor}
\usepackage{tabularx}
\usepackage{multirow}

\newcommand{\comment}[1]{{}}

\newcommand{\qed}{\hfill \ensuremath{\Box}}
\newcommand{\tr}{\mathrm{Tr}}

\begin{document}

\preprint{APS/123-QED}

\title{Detecting unfaithful entanglement by multiple fidelities}

 \author{Ruiqi Zhang$^{1,2}$}
 \author{Zhaohui Wei$^{1,3,}$}\email{Email: weizhaohui@gmail.com}
 \affiliation{$^{1}$Yau Mathematical Sciences Center, Tsinghua University, Beijing 100084, China\\$^{2}$Department of Mathematics, Tsinghua University, Beijing 100084, China\\$^{3}$Yanqi Lake Beijing Institute of Mathematical Sciences and Applications, 101407, China}

\begin{abstract}
Certifying entanglement {for unknown quantum states} experimentally is a fundamental problem in quantum computing and quantum physics. {Because of} being easy to implement, a most popular approach for this problem in modern quantum experiments is detecting target quantum states with fidelity-based entanglement witnesses. Specifically, if the fidelity between a target state and an entangled pure state exceeds a certain value, the target state can be guaranteed to be entangled. Recently, however, it has been realized that there exist {so-called unfaithful quantum states, which can be entangled, but} their entanglement cannot be certified by any fidelity-based entanglement witnesses. In this paper, by specific examples we show that if one makes a slight modification to fidelity-based entanglement witnesses by combining multiple fidelities together, it is still possible to certify entanglement for unfaithful quantum states with this popular technique. Particularly, we will analyze the mathematical structure of the modified entanglement witnesses, and propose an algorithm that can search for the optimal designs for them.
\end{abstract}

\maketitle

\section{Introduction}\label{introduction}

Characterizing unknown quantum systems in various experimental environments is a most fundamental problem in quantum computing and quantum physics~\cite{brydges2019probing,huang2020predicting,elben2020mixed,toth2005entanglement,dimic2018single,walborn2006experimental,friis2019entanglement,lin2023quantifying,chen2023certifying,cao2023generation,asif2023entanglement,lin2023all}. For an unknown quantum system, when the full information of its state needs to be found out, one could use the technique of quantum state tomography (QST) to obtain the density matrix~\cite{chuang1997prescription, poyatos1997complete,thew2002qudit,cramer2010efficient,torlai2018neural}. Usually, in a QST task one has to measure the target quantum system in many different measurement settings and then analyze the collected outcome statistics. Although in principle all the relevant quantum properties can be found out after the density matrix is known, the experimental cost of QST is extremely high.

Meanwhile, sometimes we are only interested in certain partial information of target quantum systems, and if this is the case, we would like to utilize much more efficient approaches than QST to characterize the target properties. For example, in experiments for quantum computing it is often the information on quantum entanglement are very valuable to us~\cite{ekert1991quantum,bennett1993teleporting,raussendorf2001one}, and in some cases we even only need to figure out whether quantum entanglement exists or not. For such tasks, a technique called \emph{entanglement witness} has been proposed to certify the existence of entanglement~\cite{horodecki1996necessary,haffner2005scalable,leibfried2005creation,guhne2009entanglement,horodecki2009quantum,lewenstein2000optimization,vznidarivc2007detecting}.

Basically, an entanglement witness is usually an operator $E$ designed in such a way that for any separable quantum states $\sigma$, it always holds that $\tr(\sigma E)\geq 0$. Therefore, if somehow we know that $\tr(\rho E)<0$ for a quantum state $\rho$, then we must have that $\rho$ is entangled.

Due to the elegant design and the fact that they are very easy to implement physically, entanglement witnesses have been widely utilized in quantum experiments to certify entanglement.
Particularly, a very popular class of entangled witnesses are the so-called fidelity-based entanglement witnesses~\cite{bourennane2004experimental,erhard2018experimental,bavaresco2018measurements}, which certify the entanglement of a quantum state $\rho$ by choosing a proper pure state $\ket{\psi}$ and then showing that the fidelity between $\rho$ and $\ket{\psi}$ is above a certain threshold.

However, recently it has been realized that in some sense the power of fidelity-based entanglement witnesses is quite limited, as there exist the so-called \emph{unfaithful} entangled quantum states whose entanglement cannot be detected by any fidelity-based entanglement witnesses~\cite{weilenmann2020entanglement}. And what is more, when quantum dimension is high it turns out that most random quantum states are unfaithful~\cite{weilenmann2020entanglement,guhne2021geometry}.

In such a situation, it is tempting to ask the following question: Since in the past decades fidelity-based entanglement witnesses have been utilized so widely, can we make some slight modification to them such that the new entanglement witnesses can detect  entanglement for unfaithful quantum states?

In this paper, we provide an affirmative answer to the above question by showing that if one designs entanglement witnesses by combining $k$ different fidelities together, which are called \emph{$k$-tuple fidelity-based entanglement witnesses}, it is still possible to certify  entanglement for unfaithful quantum states with this popular techniques. More specifically, we first provide examples that 2-tuple fidelity information can certify unfaithful entanglement state, which demonstrates the advantages of multiple fidelities over original fidelity-based entanglement witnesses. Secondly, we show that the experience of selecting maximally entangled states to design original fidelity-based entanglement witnesses, for which the optimality has been proved~\cite{guff2022optimal}, often fails in the design of 2-tuple fidelity-based entanglement witnesses. Finally, based on variational generative optimization network (VGON)~\cite{zhang2024variational}, a new approach to solve optimization problems with generative models in machine learning, we propose an algorithm to search for the optimal $k$-tuple fidelity-based entanglement witnesses. Lastly we perform numerical calculations on several nontrivial examples to demonstrate the good performance of our algorithm in designing entanglement witnesses.

The rest of this paper is organized as follows. Section II introduces the background of the problems we aim to solve. Section III reexamines the problem that quantum noise transfers pure quantum states to unfaithful quantum states, which will be useful in our later discussions. Section IV presents a specific example that entanglement for unfaithful states can still be certified if one uses 2-tuple fidelity-based entanglement witnesses. In Section V, we show that the maximally entangled states commonly chosen in the design of original fidelity-based entanglement witnesses perform poorly when designing 2-tuple fidelity-based entanglement witnesses. Section VI propose an algorithm to design optimal $k$-tuple fidelity-based entanglement witnesses, and finally in Section VII we numerically demonstrates that the algorithm has a good performance in searching for high-quality entanglement witnesses. 

\section{Preliminaries}

In this paper, we mainly focus on bipartite quantum states. Suppose $\rho$ is a quantum state shared by two parties, Alice and Bob, then we say $\rho$ is a separable state if it can be expressed as a convex combination (probability mixture) of states of individual subsystems:
$$
\rho = \sum_i p_i \, \rho_i^A \otimes \rho_i^B,
$$
where $\rho_i^A$ and $\rho_i^B$ are density matrices of subsystems A and B, respectively, and $\{p_i\}$ is a probability distribution such that $p_i \geq 0$ and $\sum_i p_i = 1$. We denote the set of all separable states by $\mathrm{SEP}$. If a bipartite quantum state is not a separable state, we say it is entangled.

For a given quantum state $\rho$, determining whether it is entangled or not is a fundamental problem. However, this problem has been shown to be NP-hard, even if the density matrix is given~\cite{gurvits2003classical}. An important relevant fact is that any separable quantum state must satisfy the so-called Positive Partial Transpose (PPT) criterion~\cite{doherty2004complete}. Specifically, for a quantum state $\rho$ the partial transpose with respect to one of the subsystems, say $B$, is obtained by transposing only the indices associated with $B$:
$$
(\rho^{T_B})_{ij,kl} = \rho_{ij,lk},
$$
where $\rho_{ij,kl}=\bra{i}\bra{j}\rho\ket{k}\ket{l}$ are the elements of the density matrix $\rho$. Then we say a quantum state $\rho$ satisfies the PPT criterion if $\rho^{T_B}$ (or equivalently $\rho^{T_A}$) has only non-negative eigenvalues. We denote the set of all PPT states by $\mathrm{PPT}$. It can be seen that $\mathrm{SEP} \subseteq \mathrm{PPT}$, though PPT states can be entangled. Meanwhile, although $\mathrm{SEP}$ is a convex set, it is not easy to characterize $\mathrm{SEP}$ mathematically. Therefore, in numerical calculations $\mathrm{PPT}$ is often utilized to approximate $\mathrm{SEP}$~ \cite{doherty2004complete}.

When the density matrix is not given, to determine whether a quantum state is entangled or not is even more challenging. In quantum experiments, a popular idea for this is as follows. Suppose $|\psi\rangle$ is a bipartite entangled pure state, and its Schmidt decomposition can be expressed as $|\psi\rangle = \sum_{i=0}^{d-1} s_i |i_A\rangle| i_B\rangle$, where $s_1 \geq s_2 \geq \cdots\geq s_{d-1}\geq0$ are the Schmidt coefficients, and $\{|i_A\rangle\}$ ($\{|i_B\rangle\}$) is an orthonormal basis for the subsystem $A$ ($B$). Then if $\rho$ is separable, we must have that~\cite{bourennane2004experimental}
\begin{equation}
\langle \psi|\rho|\psi\rangle \leq s_1^2.
\end{equation}
Based on this fact, one can define an entanglement witness of form
\begin{equation}
E=s_1^2I-\ket{\psi}\bra{\psi}
\end{equation}
such that a target quantum state $\rho$ can be certified to be entangled as long as $\tr(\rho E)<0$, where $I$ is the identity operator. Moreover, the scalar $s_1^2$ in $E$ is the optimal in that for any $a < s_1^2$, there exists a separable state $\sigma$ such that $\langle \psi|\sigma|\psi\rangle > a$.

We call the entanglement witness $E$ defined above a \emph{fidelity-based entanglement witness}. The quantum states that can be proven entangled by fidelity-based entanglement witnesses are called \emph{faithful states}. Because of being easy to physically implement, fidelity-based entanglement witnesses are widely applied in quantum experiments to certify the existence of quantum entanglement.

Recently, it has been realized that it is possible that a quantum state $\rho$ is entangled, but it satisfies that $\langle \phi|\rho|\phi\rangle \leq t_1^2$ for any pure state $|\phi\rangle$ with Schmidt decomposition $|\phi\rangle = \sum_i t_i |i_A \rangle|i_B\rangle$, which means that this type of quantum entanglement cannot be certified by any fidelity-based entanglement witnesses. When this is the case, $\rho$ is called an \emph{unfaithful state}. We denote the set of all unfaithful states as $U_2$. 

It turns out that the set $U_2$ is also very challenging to characterize, and one can use another set $\widetilde{U}_2$, defined by the following semidefinite program (SDP), to approximate $U_2$ from the inside, i.e., $\widetilde{U}_2 \subseteq U_2$~\cite{weilenmann2020entanglement, guhne2021geometry}.\\

\noindent\textbf{SDP 1}. Let $\rho_{A B}$ be a bipartite state. If there exists $\mu \in$ $[0,1]$ and positive semidefinite operators $M_A, M_B$ such that
$$
\begin{aligned}
& M_A \otimes \mathbb{I}_B+\mathbb{I}_A \otimes M_B \geq \rho_{A B}, \\
& \mu(D-1)=\operatorname{tr}\left[M_A\right], \quad \mu \mathbb{I}-M_A \geq 0, \\
& (1-\mu)(D-1)=\operatorname{tr}\left[M_B\right], \quad(1-\mu) \mathbb{I}-M_B \geq 0,
\end{aligned}
$$
then we say that $\rho_{A B} \in \widetilde{U}_D$, where $D$ is an integer.

\section{Quantum Noise Leads to Unfaithfulness}

It can be seen that fidelity-based entanglement witnesses fail to certify entanglement only for mixed quantum states. In quantum experiments, the mixedness of quantum states often comes from quantum noise. In other words, under the impact of quantum noise, certifying entanglement becomes a challenging task for fidelity-based entanglement witnesses. In this paper, we mainly consider two types of quantum noise: the depolarizing noise and the dephasing noise. For a bipartite quantum state $\rho$ with a local dimension $d$, the depolarizing noise transfers $\rho$ to
$$
\tilde{\rho}(p) = p\frac{I_{d^2}}{d^2}+(1-p)\rho,
$$
where $p \in [0,1]$ is the strength of the noise, and the dephasing noise transfers $\rho$ to
$$
\tilde{\rho}(p) =  p\mathcal{D}(\rho)+(1-p)\rho,
$$
where $p \in [0,1]$ is again the strength of the noise and $\mathcal{D}(\cdot)$ is an operation that keeps the diagonal entries of $\rho$ unchanged and sets all the off-diagonal entries to be $0$.

In a typical quantum experiment, usually it is an entangled pure state $\rho$ that is prepared \cite{leibfried2005creation, walborn2006experimental}. However, due to the existence of quantum noise, $\rho$ might be transferred to an unfaithful state $\tilde{\rho}(p)$ as long as the strength of noise is strong enough. This kind of phenomenon was first studied in \cite{weilenmann2020entanglement}. For the convenience of later discussions, in this section we will reexamine this topic. Particularly, we will quantitatively analyze the level of noise that can cause unfaithfulness.

\subsection{The case of pure quantum states}

We suppose that a pure quantum state $\rho=\ket{\psi}\bra{\psi}$ suffers from a global depolarizing noise of strength $p$, and $p$ increases from $0$ to $1$ continuously. We would like to find out when $\tilde{\rho}(p) = p \frac{I_{d^2}}{d^2} + (1-p) \rho$ becomes unfaithful, and when it becomes separable.\\

\noindent\textbf{Theorem 3.1.} $|\psi\rangle$ is a bipartite pure quantum state, $|\psi\rangle\in \mathcal{H}_A\otimes \mathcal{H}_B$, and $\dim \mathcal{H}_A = \dim \mathcal{H}_A = d$. Assume that its Schmidt decomposition can be written as $|\psi\rangle = \sum_{i=1}^d s_i |i_A\rangle|i_B\rangle$, where $s_1\geq s_2 \geq \cdots s_d \geq 0$, and $\{|i_A\rangle\}_{i=1}^d$ and $\{|i_A\rangle\}_{i=1}^d$ are standard orthonormal bases of the spaces $\mathcal{H}_A$ and $\mathcal{H}_B$, respectively. Then the quantum state $\tilde\rho(p) = p I /d^2 + (1-p)|\psi\rangle\langle\psi|$ is unfaithful if and only if
\begin{equation}
\frac{d\left(\sum_{i} s_i\right)^2 - d}{d\left(\sum_{i} s_i\right)^2 -1}\leq p\leq 1.
\end{equation}

\noindent\textbf{Proof:} Suppose $\varrho_{A B}$ is a bipartite quantum state. Then $\varrho_{A B}$ is faithful if and only if there are local unitary transformations $U_A$ and $U_B$ such that~\cite{guhne2021geometry}
$$
\left\langle\phi^{+}\left|U_A \otimes U_B \varrho_{A B} U_A^{\dagger} \otimes U_B^{\dagger}\right| \phi^{+}\right\rangle>\frac{1}{d},
$$
where $| \phi^{+}\rangle=\sum_{i=1}^d \frac{1}{\sqrt{d}} |i\rangle|i\rangle$ is a maximally entangled state.

Since $\rho=\ket{\psi}\bra{\psi}$, it holds that $\tilde\rho(p) = p I /d^2 + (1-p)|\psi\rangle\langle\psi|$. Let $|\psi\rangle = \sum_{i=1}^d s_i |i_A\rangle|i_B\rangle$ is a Schmidt decomposition, where $s_1\geq s_2 \geq \cdots s_d \geq 0$, and $\{|i_A\rangle\}_{i=1}^d$ and $\{|i_B\rangle\}_{i=1}^d$ are standard orthonormal bases of the spaces $\mathcal{H}_A$ and $\mathcal{H}_B$, respectively. Assume that $|\psi_2\rangle = \sum_{i=1}^d s_i |i\rangle|i\rangle$. Then
$$
\begin{aligned}
 & \max_{U_A, U_B} \langle\phi^{+}|U_A \otimes U_B \tilde\rho(p) U_A^{\dagger} \otimes U_B^{\dagger}| \phi^{+}\rangle\\
 = & \max_{U_A, U_B} \langle\phi^{+}|U_A \otimes U_B (p I /d^2 + (1-p)|\psi\rangle\langle\psi|) U_A^{\dagger} \otimes U_B^{\dagger}| \phi^{+}\rangle\\
 = & \max_{U_A, U_B} \langle\phi^{+}|U_A \otimes U_B (p I /d^2 + (1-p)|\psi_2\rangle\langle\psi_2|) U_A^{\dagger} \otimes U_B^{\dagger}| \phi^{+}\rangle\\
 = & \max_{U_A, U_B} \langle\phi^{+}|I \otimes U_A^T U_B (p I /d^2 + (1-p)|\psi_2\rangle\langle\psi_2|) I^{\dagger} \otimes (U_A^T U_B)^{\dagger}| \phi^{+}\rangle.\\
\end{aligned}
$$
It can be verified that the optimization problem attains its maximum value when $U_A^T U_B=I$. Then we have that
$$
\max_{U_A, U_B} \langle\phi^{+}|U_A \otimes U_B \tilde\rho(p) U_A^{\dagger} \otimes U_B^{\dagger}| \phi^{+}\rangle = \frac{p}{d^2} + (1-p)\frac1d\left(\sum_{i=1}^ds_i\right)^2.
$$
Therefore we obtain that
when $\frac{p}{d^2} + (1-p)\frac1d(\sum_{i=1}^ds_i)^2\leq 1/d$ , $\tilde\rho(p)$ is unfaithful. When $\frac{p}{d^2} + (1-p)\frac1d(\sum_{i=1}^ds_i)^2> 1/d$ , $\tilde\rho(p)$ is faithful. After simplification, the proof is concluded.

\qed

From Theorem 3.1 we can see that for the pure state $\rho=\ket{\psi}\bra{\psi}$, there is a threshold such that when the noise level is larger than the threshold, $\tilde\rho(p)$ becomes unfaithful. Since $U_2$, the set of unfaithful quantum states, is a convex set, this kind of threshold also exists for any mixed faithful state $\sigma$, which we denote $P_{U_2}(\sigma)$. Theorem 3.1 shows that $P_{U_2}(\rho)=\frac{d\left(\sum_{i} s_i\right)^2 - d}{d\left(\sum_{i} s_i\right)^2 -1}$.

Similarly, for an entangled state $\tau$, there also exists a threshold, denoted $P_{\mathrm{SEP}}(\tau)$, such that $\tilde{\tau}(p)= p I /d^2 + (1-p)\tau$ is separable when $P_{\mathrm{SEP}}(\tau)\leq p\leq1$ and entangled otherwise. The following lemma give us the value of $P_{\mathrm{SEP}}(\rho)$, which was first reported in Ref.\cite{vidal1999robustness}.\\

\noindent\textbf{Lemma 2.} Suppose $|\psi\rangle\in \mathcal{H}_A\otimes \mathcal{H}_B$ is defined as in Theorem 3.1. Then the quantum state $\tilde\rho(p) = p I /d^2 + (1-p)|\psi\rangle\langle\psi|$ satisfies the PPT criterion if and only if it is separable, where $p\in[0,1]$. Furthermore, it holds that
\begin{equation}
P_{\mathrm{SEP}}(\rho) = \frac{d^2 s_1s_2}{1+d^2 s_1s_2},
\end{equation}
implying that $\tilde\rho(p)$ is separable if and only if $P_{\mathrm{SEP}}(\rho)\leq p\leq1$.

It can be verified that for any pure state $\rho$, we always have $P_{\mathrm{SEP}}(\rho) \geq P_{U_2}(\rho)$. Moreover, if and only if $\rho$ is maximally entangled or the subsystem dimension $d = 2$, it holds that $P_{\mathrm{SEP}}(\rho) = P_{U_2}(\rho)$. Therefore, for a typical pure state $\rho$, when $P_{U_2}(\rho)<p<P_{\mathrm{SEP}}(\rho)$, $\tilde\rho(p)$ is unfaithful and entangled, whose entanglement cannot be detected by any fidelity-based entanglement witnesses.

\subsection{The case of mixed quantum states}

We next move to the case that the ideal target quantum state $\rho$ is mixed and entangled. It turns out that this case is hard to handle in an analytical way, so we will numerically show that for a random mixed faithful and entangled state $\rho$, quantum noise also transfer it to an unfaithful and entangled state when the strength of quantum noise $p$ is in a certain interval (see Fig. 1). As a result, fidelity-based entanglement witnesses also fail in detecting the underlying entanglement.

Similar to the case of pure states, we now show that $P_{\mathrm{SEP}}(\rho)> P_{U_2}(\rho)$ for most mixed states $\rho$. However, when $\rho$ is mixed, both $P_{\mathrm{SEP}}(\rho)$ and $P_{U_2}(\rho)$ are hard to calculate analytically. To overcome this difficulty, we consider using $\mathrm{PPT}$ to approximate $\mathrm{SEP}$ from the outside, and using $\widetilde{U}_2$ to approximate $U_2$ from the inside. This lead us to define the quantity $P_{\mathrm{SEP}}^{\mathrm{inf}}(\rho)$ such that $p \geq P_{\mathrm{SEP}}^{\mathrm{inf}}(\rho)$ if and only if $\tilde{\rho}(p)$ is a PPT state. Similarly, we define the quantity $P_{U_2}^{\mathrm{sup}}(\rho)$ such that $p \geq P_{U_2}^{\mathrm{sup}}(\rho)$ if and only if $\tilde{\rho}(p)$ satisfies SDP 1 for $D = 2$, i.e., $\tilde{\rho}(p)\in \widetilde{U}_2$.
Since $\mathrm{SEP} \subseteq \mathrm{PPT}$, it holds that $P_{\mathrm{SEP}}(\rho) \geq P_{\mathrm{SEP}}^{\mathrm{inf}}(\rho)$. Since $\widetilde{U}_2 \subseteq U_2$, we also have $P_{U_2}^{\mathrm{sup}}(\rho) \geq P_{U_2}(\rho)$. Therefore, one can show $P_{\mathrm{SEP}}(\rho)> P_{U_2}(\rho)$ by numerically showing that $P_{\mathrm{SEP}}^{\mathrm{inf}}(\rho) > P_{U_2}^{\mathrm{sup}}(\rho)$.

In fact, the values of $P_{\mathrm{SEP}}^{\mathrm{inf}}(\rho)$ and $P_{U_2}^{\mathrm{sup}}(\rho)$ can be calculated using the following SDP:
\begin{equation}\label{optP1-1}
\begin{array}{ll}
\min & p \\
\text {subject to } & \tilde\rho(p)  \in \mathrm{D}, \\
& p\in [0,1],
\end{array}
\tag{P1}
\end{equation}
where $\tilde\rho(p) = p\frac{I_{d^2}}{d^2}+(1-p)\rho$, and $D$ is set to be $\mathrm{PPT}$ and $\widetilde{U}_2$ respectively. Note that the sets $\mathrm{PPT}$ and $\widetilde{U}_2$ can be characterized by SDPs, so the above optimization problems are indeed SDPs.

Recall that the above discussions suppose that all the quantum noise is the depolarizing noise. When the underlying noise is the dephasing noise, one simply needs to replace the relation $\tilde{\rho}(p) = p \frac{I_{d^2}}{d^2} + (1-p)\rho$ with $\tilde{\rho}(p) =  p\mathcal{D}(\rho)+(1-p)\rho$, and the remaining analyses are similar.

In our numerical calculations, we randomly select 100 quantum states according to the Haar distribution with the subsystem dimension $d = 2, 3, 4$. Then we calculate the above boundary points for the sets $\mathrm{PPT}$ and $\tilde U_2$, where both the depolarizing noise and the dephasing noise are considered. In turns out that on all the picked quantum states we observe that $P_{\mathrm{SEP}}^{\mathrm{inf}}(\rho) > P_{U_2}^{\mathrm{sup}}(\rho)$. The results are illustrated in Fig.1.

\begin{figure}[!ht]
    \centering
    \includegraphics[width=0.48\textwidth]{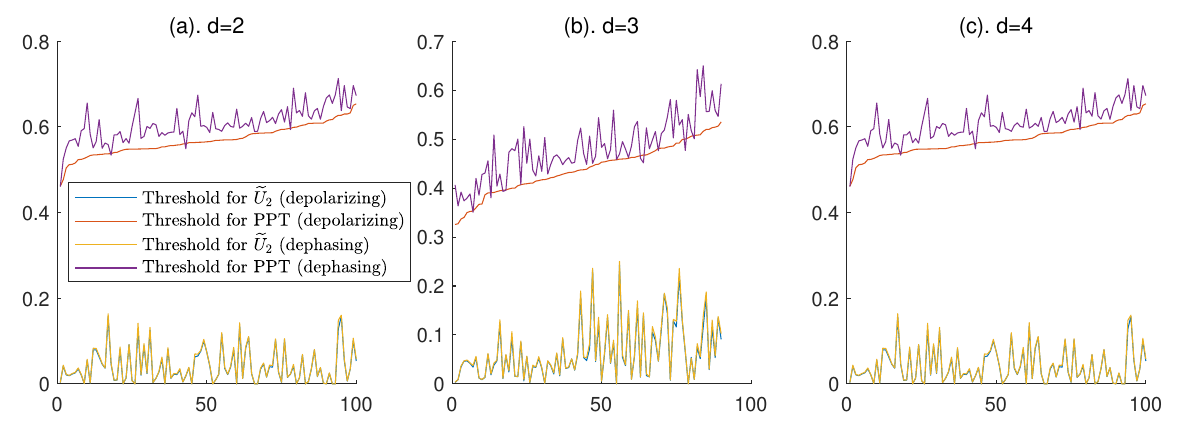}
    \caption{We randomly select faithful quantum states $\rho$ according to Haar or Bures distribution with the local dimension $d = 2,3,4$. We then solve Problem (P1) to obtain the thresholds of the noise strength $p$ for the depolarizing noise or the dephasing noise that make the quantum states in $\widetilde{U}_2$ (inner approximation of the unfaithful set $U_2$) or PPT. All of the faithful states we select become unfaithful and entangled states when the noise strengths are in certain intervals.}
    \label{fig:result_General}
\end{figure}

\section{Measuring two different fidelities can detect unfaithful  entanglement}

According to the discussion in the previous section, we know that when the strength of noise $p$ is in the interval $[P_{U_2}(\rho), P_{\mathrm{SEP}}(\rho))$, quantum noise transfers the target quantum state $\rho$ to an unfaithfully entangled state $\tilde\rho(p)$, making any fidelity-based entanglement witness fail to detect the underlying entanglement. In such a situation, a natural question arises: Since the implementations of fidelity-based entanglement witnesses are experiment-friendly, can we make a slight modification to them such that the entanglement of $\tilde{\rho}(p)$ can be detected even if $p\in[P_{U_2}(\rho), P_{\mathrm{SEP}}(\rho))$?

Recall that the basic idea of fidelity-based entanglement witness is that, a quantum state $\rho$ is proven to be entangled, if there exists a pure state $|\psi\rangle$ such that for any $\sigma \in \mathrm{SEP}$ it holds that $\langle \psi | \rho | \psi \rangle \neq \langle \psi | \sigma | \psi \rangle$. Then we can generalize fidelity-based entanglement witness as below. For a quantum state $\rho$, if one can find two pure quantum states $|\psi_1\rangle$ and $|\psi_2\rangle$ such that for any $\sigma \in \mathrm{SEP}$, $\langle \psi_1 | \rho | \psi_1 \rangle = \langle \psi_1 | \sigma | \psi_1 \rangle$ and $\langle \psi_2 | \rho | \psi_2 \rangle = \langle \psi_2 | \sigma | \psi_2 \rangle$ cannot hold at the same time, then $\rho$ can be proven to be entangled. If this is the case, we say that the entanglement of $\rho$ is certified by a \emph{2-tuple fidelity-based entanglement witness}. 

For the convenience of later discussions, we define
\begin{equation}\label{eq:w2SEP}
\begin{aligned}& W_2(|\psi_1\rangle, |\psi_2\rangle) \\
= & \left\{ \rho: \exists \sigma \in \mathrm{SEP} \text{ such that } \langle\psi_i|\rho|\psi_i\rangle = \langle\psi_i|\sigma|\psi_i\rangle, i = 1,2 \right\}.
\end{aligned}
\end{equation}
Then if one finds out that $\rho \notin W_2(|\psi_1\rangle, |\psi_2\rangle)$, $\rho$ must be entangled. Since $\mathrm{SEP}$ is hard to characterize numerically and can be approximated by $\mathrm{PPT}$, we define an outer approximation of $W_2(|\psi_1\rangle, |\psi_2\rangle)$, i.e.,
\begin{equation}\label{eq:w2PPT}
\begin{aligned}&\tilde W_2(|\psi_1\rangle, |\psi_2\rangle) \\
  = & \left\{ \rho: \exists \sigma \in \mathrm{PPT} \text{ such that } \langle\psi_i|\rho|\psi_i\rangle = \langle\psi_i|\sigma|\psi_i\rangle, i = 1,2 \right\}.
  \end{aligned}
\end{equation}
It can be seen that $W_2(|\psi_1\rangle, |\psi_2\rangle) \subseteq \tilde W_2(|\psi_1\rangle, |\psi_2\rangle)$, and therefore if $\rho \notin \tilde W_2(|\psi_1\rangle, |\psi_2\rangle)$, we will have that $\rho$ is entangled. Moreover, whether $\rho \in \tilde W_2(|\psi_1\rangle, |\psi_2\rangle)$ or not can be numerically determined by SDP.

We now exhibit specific examples showing that 2-tuple fidelity-based entanglement witnesses indeed can detect unfaithful entanglement. Consider quantum states of form
\begin{equation}\label{eq:rank2}
\rho = q_1 |\phi_1\rangle\langle\phi_1| + q_2 |\phi_2\rangle\langle\phi_2|,
\end{equation}
where $|\phi_1\rangle = \frac{1}{\sqrt{d}}\sum_{i=1}^{d} |ii\rangle$, $|\phi_2\rangle = |12\rangle$, $q_1+q_2=1$, and $q_1\in \{0.1, 0.2, \cdots, 1.0\}$.

To implement the above modified version of fidelity-based entanglement witness, we choose $|\psi_i\rangle = |\phi_i\rangle$ for $i = 1, 2$. Then by measuring the target state, we obtain the fidelity $\langle\psi_i|\rho|\psi_i\rangle$ for $i=1,2$. Next by SDP, we calculate the minimum value of $p$ such that $\tilde{\rho}(p) \in \tilde W_2(|\psi_1\rangle, |\psi_2\rangle)$, and we still denote this value by $\mathcal{F}(\psi_1,\psi_2)$. This means that when $p < \mathcal{F}(\psi_1,\psi_2)$, $2$-tuple fidelity-based entanglement witnesses can certify that $\tilde{\rho}(p)$ is entangled.

For example, if we set $q_1=0.1$ in Eq. \eqref{eq:rank2}, we can obtain that $\mathcal{F}(\psi_1,\psi_2) = 0.122097$, and $P_{U_2}^{\mathrm{sup}}(\rho)$ and $P_{\mathrm{SEP}}^{\mathrm{inf}}(\rho)$  are $0.016448$ and $0.285721$ respectively. This implies that when $0.016448<p<0.285721$, $\tilde{\rho}(p)$ is an unfaithfully entangled quantum state, meaning that original fidelity-based entanglement witnesses fail to detect its entanglement. However, when $0.016448<p<0.122097$, although the underlying entanglement is unfaithful, the 2-tuple fidelity-based entanglement witness we propose above can successfully certify the entanglement. Therefore, it can be seen that by slightly increasing the experimental cost, one can significantly improves the power of fidelity-based entanglement witnesses, i.e., they can certify entanglement for unfaithful quantum states. More results can be seen in TABLE \ref{table:result_diffdim1}.

\begin{table} \scriptsize
\caption{The entanglement for the quantum states $\rho = q_1 |\phi_1\rangle \langle \phi_1| + q_2 |\phi_2\rangle \langle \phi_2|$ defined in Eq. \eqref{eq:rank2}. To construct 2-tuple fidelity-based entanglement witnesses, we choose $|\psi_i\rangle = |\phi_i\rangle$ for $i = 1, 2$. The Table provides the thresholds of $p$ for $\tilde{\rho}(p)$ to be in PPT, in $\widetilde{U}_2$, and to be detected as entangled by 2-tuple fidelity-based entanglement witnesses. Specifically, when $p < P_{\mathrm{SEP}}^{\mathrm{inf}}(\rho)$, $\tilde{\rho}(p)$ is entangled; when $p \geq P_{U_2}^{\mathrm{sup}}(\rho)$, $\tilde{\rho}(p)$ is unfaithful; and when $p < \mathcal{F}(\psi_1, \psi_2)$, the 2-tuple fidelity-based entanglement witnesses involving $|\psi_1\rangle, |\psi_2\rangle$ can certify that $\tilde{\rho}(p)$ is entangled. Notice that $\mathcal{F}(\psi_1, \psi_2) > P_{U_2}^{\mathrm{sup}}(\rho)$, implying that 2-tuple fidelity-based entanglement witnesses can certify unfaithful entanglement.}
\setlength{\tabcolsep}{3mm}{
\begin{tabular}{cccc}
\hline
\hline $q_1$ value & $P_{\mathrm{SEP}}^{\mathrm{inf}}(\rho)$ & $P_{U_2}^{\mathrm{sup}}(\rho)$ & $\mathcal{F}\left(\psi_1, \psi_2\right)$ \\
\hline 0.1 & 0.28572053 & 0.016448152 & 0.12209736 \\
 0.2 & 0.44444444 & 0.081603629 & 0.24170884 \\
 0.3 & 0.54546258 & 0.23076806 & 0.35773928 \\
 0.4 & 0.61536874 & 0.44444469 & 0.46848499 \\
 0.5 & 0.66666745 & 0.57142865 & 0.57142883 \\
 0.6 & 0.70588197 & 0.65115927 & 0.65115477 \\
 0.7 & 0.73684672 & 0.70588378 & 0.70588223 \\
 0.8 & 0.76191575 & 0.74576031 & 0.74576686 \\
 0.9 & 0.78261339 & 0.77613138 & 0.77611945 \\
\hline
\hline
\end{tabular}}
\label{table:result_diffdim1}
\end{table}

However, in order to design $2$-tuple fidelity-based entanglement witnesses, how to choose proper $|\psi_1\rangle$ and $|\psi_2\rangle$ is still an apparent issue. In the next two sections, we will address this problem.

\section{Maximally entangled states may not good choices to construct 2-tuple fidelity-based entanglement witnesses}

It has been known that the optimal constructions of fidelity-based entanglement witnesses can always be based on maximally entangled states, i.e., such a witness can be chosen as $E=\frac{1}{d}\cdot I-\ket{\psi}\bra{\psi}$, where $d$ is the local dimension, and $\ket{\psi}$ is a bipartite maximally entangled state~\cite{guhne2021geometry,guff2022optimal}. However, we now show that if both $|\psi_1\rangle$ and $|\psi_2\rangle$ of a 2-tuple fidelity-based entanglement witness are chosen as maximally entangled states, it will probably not have any advantage over original fidelity-based entanglement witnesses.

Without loss of generality, we suppose
\begin{equation}\label{eq:maximal1}
|\psi_1\rangle = \frac{1}{\sqrt{d}} \sum_{i=1}^d |i_A\rangle\otimes|i_B\rangle, \ \ |\psi_2\rangle = \frac{1}{\sqrt{d}} \sum_{i=1}^d |i_A\rangle\otimes U |i_B\rangle,
\end{equation}
where $U$ is a local unitary acting on Bob's subsystem. we will show that if $\rho \notin W_2(|\psi_1\rangle,|\psi_2\rangle)$, then $\rho \notin U_2$. This means that if a 2-tuple fidelity-based entanglement witness using $|\psi_1\rangle, |\psi_2\rangle$ can detect the entanglement of $\rho$, then some original fidelity-based entanglement witness can also achieve this. Therefore, the experience of choosing maximally entangled states to design optimal fidelity-based entanglement witnesses is not correct any more in the new setting. 

To prove the above claim, we define the following function that maps a quantum state to a 2-tuple vector:
\begin{equation}
f_{|\psi_1\rangle,|\psi_2\rangle}(\rho): \rho \mapsto (\langle\psi_1|\rho|\psi_1\rangle, \langle\psi_2|\rho|\psi_2\rangle).
\end{equation}
Then for any given $|\psi_1\rangle, |\psi_2\rangle$, if $f_{|\psi_1\rangle,|\psi_2\rangle}(\rho) \notin f_{|\psi_1\rangle,|\psi_2\rangle}(\mathrm{SEP})$, $\rho$ is entangled, where we have applied the function $f$ onto sets of quantum states. Similarly, if $f_{|\psi_1\rangle,|\psi_2\rangle}(\rho) \notin f_{|\psi_1\rangle,|\psi_2\rangle}(U_2)$, $\rho$ is a faithful state. Before proceeding, let us introduce two facts on $f_{|\psi_1\rangle,|\psi_2\rangle}(\rho)$ when the input $\rho$ is separable states. \\

\noindent\textbf{Lemma 5.1}: For $|\psi_1\rangle, |\psi_2\rangle$ defined in Eq. \eqref{eq:maximal1}, there always exists a separable state $\rho$ such that $|\langle\psi_i|\rho|\psi_i\rangle| = 1/d, i = 1,2$.\\

\noindent\textbf{Proof:} 
We will show that there exists a pure separable state $\rho$ that can achieve this. For this, we suppose $\rho = |\phi\rangle\langle\phi| = |x_A\rangle\langle x_A| \otimes |x_B\rangle\langle x_B|$. In order to make $\left|\langle\psi_j|\rho|\psi_j\rangle\right| = 1/d, j = 1,2$ hold, we need $|x_A\rangle, |x_B\rangle$ to satisfy
$$
\left|\sum_{i = 1}^d\langle i_A|x_A\rangle\langle i_B|x_B\rangle\right|^2 = 1,\left|\sum_{i = 1}^d\langle i_A|x_A\rangle\langle i_B|U^\dagger |x_B\rangle\right|^2 = 1.
$$
Since $\exists |y_A\rangle$ such that $\forall i, \langle i_A|x_A\rangle = \langle y_A| i_B\rangle$. Therefore, the above two equations can be simplified to
$$
\left|\sum_{i = 1}^d\langle y_A|i_B\rangle\langle i_B|x_B\rangle\right|^2 = 1,\left|\sum_{i = 1}^d\langle y_A|i_B\rangle\langle i_B|U^\dagger |x_B\rangle\right|^2 = 1.
$$
They are equal to
$$
\left|\langle y_A|x_B\rangle\right|^2 = 1,\left|\langle y_A|U^\dagger |x_B\rangle\right|^2 = 1.
$$
Since $\left|\langle y_A|x_B\rangle\right|^2 = 1$ we know that $|y_A\rangle$ and $|x_B\rangle$ differ by at most a global phase, thus the two equations hold if and only if there exists an $|x_B\rangle$ such that $\left|\langle x_B|U^\dagger |x_B\rangle\right|^2 = 1$. Since $U$ is a complex normal operator and the norms of the eigenvalues are all $1$, such a $|x_B\rangle$ always exists.

\qed

\noindent\textbf{Lemma 5.2}: Suppose $|\psi_1\rangle, |\psi_2\rangle$ are defined in Eq. \eqref{eq:maximal1}. Then there exists a separable state $\rho$ such that $|\langle\psi_1|\rho|\psi_1\rangle| = 1/d$ and $|\langle\psi_2|\rho|\psi_2\rangle| = 0$ if and only if one can find a vector $|x\rangle$ such that $\langle x|U|x\rangle = 0$, where $U$ is given in Eq. \eqref{eq:maximal1}.\\

\noindent\textbf{Proof:} 
As in the previous proof, we can still assume that $\rho$ is a pure state and can be written as $\rho = |\phi\rangle\langle\phi| = |x_A\rangle\langle x_A| \otimes |x_B\rangle\langle x_B|$. From $|\psi_1\rangle = \frac 1{\sqrt{d}} \sum_{i=1}^d |i_A\rangle|i_B\rangle$ and $|\psi_2\rangle = \frac 1{\sqrt{d}} \sum_{i=1}^d |i_A\rangle\otimes U |i_B\rangle$ we can get $\left|\langle\psi_1|\rho|\psi_1\rangle\right| = 1/d, \left|\langle\psi_2|\rho|\psi_2\rangle\right| = 0$ equal to
$$
\left|\sum_{i = 1}^d\langle i_A|x_A\rangle\langle i_B|x_B\rangle\right|^2 = 1,\left|\sum_{i = 1}^d\langle i_A|x_A\rangle\langle i_B|U^\dagger |x_B\rangle\right|^2 = 0.
$$
Since $\exists |y_A\rangle$ such that $\forall i, \langle i_A|x_A\rangle = \langle y_A| i_B\rangle$. Therefore, the above two equations can be simplified to
$$
\left|\sum_{i = 1}^d\langle y_A|i_B\rangle\langle i_B|x_B\rangle\right|^2 = 1,\left|\sum_{i = 1}^d\langle y_A|i_B\rangle\langle i_B|U^\dagger |x_B\rangle\right|^2 = 0,
$$
and they are equal to
$$
\left|\langle y_A|x_B\rangle\right|^2 = 1,\left|\langle y_A|U^\dagger |x_B\rangle\right|^2 = 0.
$$
Again, from $\left|\langle y_A|x_B\rangle\right|^2 = 1$ we know that $|y_A\rangle$ and $|x_B\rangle$ are the same up to a global phase, thus the two equations hold if and only if there exists an $|x_B\rangle$ such that $\left|\langle x_B|U^\dagger |x_B\rangle\right|^2 = 0$.

\qed
Note that one can find a unitary $U$ such that for any pure state $|x\rangle$ it holds that $\left|\langle x|U |x\rangle\right|^2 \neq 0$. An example for such a $U$ is the identity operator. However, we would like to stress that for many typical pairs of maximally entangled states, $U$ does satisfy the conditions in Lemma 5.2. For example, if $|\psi_1\rangle$ and $|\psi_2\rangle$ are two orthogonal maximally entangled states, then there must be a vector $|x\rangle \in \mathbb{C}^d$ such that $\langle x|U|x\rangle = 0$, which can be explained as below.

When $|\psi_1\rangle, |\psi_2\rangle$ are orthogonal, it can be verified that $\operatorname{tr}(U) = 0$. Since $U$ is unitary, then there exists a invertible matrix $Q$ such that
$$
Q^{\dagger} U Q=\left(\begin{array}{lll}
\lambda_1 & & \\
& \ddots & \\
& & \lambda_d
\end{array}\right).
$$
The fact $\operatorname{tr}(U) = 0$ indicates that $\lambda_1 + \cdots + \lambda_d = 0$. Therefore, if we take $|y\rangle = \frac{1}{\sqrt{d}}(1,1,\cdots,1)^T$ and let $|x\rangle = Q|y\rangle$, then we have $\langle x|U|x\rangle = 0$.

For an arbitrary unitary $U$, determine whether there exists a state $\ket{x}$ such that $\langle x|U|x\rangle = 0$ can be figured out by solving an linear programming problem.

We now are ready to prove the claim made at the beginning of this section, that is, if we design a 2-tuple fidelity-based entanglement witness using the quantum states $|\psi_1\rangle, |\psi_2\rangle$ in Eq. \eqref{eq:maximal1}, and if there exists a vector $|x\rangle$ such that $\langle x|U|x\rangle = 0$, then the new entanglement witness cannot detect any unfaithful entanglement.

In fact, according to the above two lemmas, when there exists a vector $|x\rangle$ such that  $\langle x|U|x\rangle = 0$, we will have that the four points $(0,0), (0,1/d),(1/d,0),(1/d,1/d) \in f_{|\psi_1\rangle,|\psi_2\rangle}(\mathrm{SEP})$. Since the set $f_{|\psi_1\rangle,|\psi_2\rangle}(\mathrm{SEP})$ is convex, we immediately have that the square $[0,1/d]^2 \subseteq f_{|\psi_1\rangle,|\psi_2\rangle}(\mathrm{SEP})$. Meanwhile, note that if $\rho$ satisfies $\langle\psi_1|\rho|\psi_1\rangle > 1/d$ or $\langle\psi_2|\rho|\psi_2\rangle > 1/d$, then $\rho$ is a faithful state~\cite{guhne2021geometry,guff2022optimal}. Therefore, for any $[x,y] \notin [0,1/d]^2$, we have $[x,y] \notin f_{|\psi_1\rangle,|\psi_2\rangle}(U_2)$, implying that $f_{|\psi_1\rangle,|\psi_2\rangle}(U_2) \subseteq [0,1/d]^2$. Combined this with the facts that $[0,1/d]^2 \subseteq f_{|\psi_1\rangle,|\psi_2\rangle}(\mathrm{SEP})$ and that $f_{|\psi_1\rangle,|\psi_2\rangle}(\mathrm{SEP}) \subseteq f_{|\psi_1\rangle,|\psi_2\rangle}(U_2)$, we have $f_{|\psi_1\rangle,|\psi_2\rangle}(\mathrm{SEP}) = f_{|\psi_1\rangle,|\psi_2\rangle}(U_2) = [0,1/d]^2$.
As a result, if both $|\psi_1\rangle$ and $|\psi_2\rangle$ are maximally entangled, and $\left|\langle x|U |x\rangle\right|^2 = 0$ for some $\ket{x}$, then the 2-tuple fidelity-based entanglement witness we design cannot detect entanglement for any unfaithfully entangled states.

\begin{figure}[!ht]
    \centering
    \includegraphics[width=0.45\textwidth]{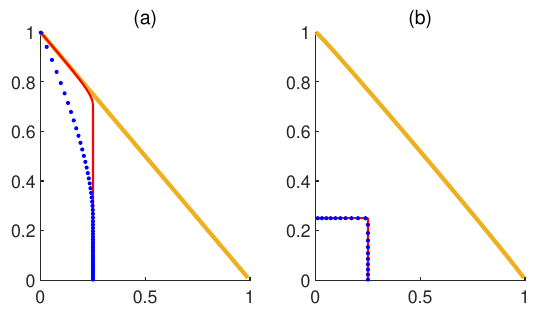}
    \caption{(a) The fidelity of quantum states with $|\psi_1\rangle = \frac{1}{\sqrt{d}}\sum_{i=1}^d|ii\rangle$, $|\psi_2\rangle = |12\rangle$ and shows the image of the function $f_{|\psi_1\rangle,|\psi_2\rangle}$. The blue point line serves as the envelope for PPT states, meaning the area below and to the left of the blue point line corresponds to $f_{|\psi_1\rangle,|\psi_2\rangle}(\text{SEP})$. The red line serves as the envelope for $\widetilde{U}_2$, meaning the area below and to the left of the red line corresponds to $f_{|\psi_1\rangle,|\psi_2\rangle}(\widetilde{U}_2)$. We notice that there is a significant gap between the red and blue point lines, which implies that if an unfaithful quantum state $\rho$ falls in the gap. We can determine the quantum state $\rho$ is entangled by our new method, i.e. 2-tuple fidelity, and fidelity-based entanglement witness cannot be detected. For (b), the only difference between (b) and (a) is that, in this figure, the quantum states $|\psi_1\rangle$ and $|\psi_2\rangle$ are chosen to be two orthogonal maximally entangled states. As a result, we find that the red line and the blue point line coincide, i.e., $f_{|\psi_1\rangle,|\psi_2\rangle}(\text{SEP}) = f_{|\psi_1\rangle,|\psi_2\rangle}(U_2) = [1, 1/d]^2$. Therefore our new methods have not any advantage compared to fidelity-based entanglement witness}
    \label{fig:figure_good.pdf}
\end{figure}

FIG. \ref{fig:figure_good.pdf} provides a intuitive explanation for this fact. In FIG. \ref{fig:figure_good.pdf}(a), we choose $|\psi_1\rangle = \frac{1}{\sqrt{d}}\sum_{i=1}^d|ii\rangle$ and $|\psi_2\rangle = |12\rangle$. The X-axis represents the fidelity between quantum states and $|\psi_1\rangle$, and Y-axis represents the fidelity between quantum states and $|\psi_2\rangle$. The blue line serves as the envelope for all the PPT states, meaning the area below and to the left of the blue point line corresponds to $f_{|\psi_1\rangle,|\psi_2\rangle}(\text{SEP})$. The red line serves as the envelope for $\widetilde{U}_2$, meaning that the area below and to the left of the orange line corresponds to $f_{|\psi_1\rangle,|\psi_2\rangle}(\widetilde{U}_2)$. The orange line connecting the coordinates $(1,0)$ and $(0,1)$ serves as the envelope for all quantum states. We notice that there is a significant gap between the red line and the blue point line, which implies that if an unfaithful quantum state $\rho$ falls in the area above and to the right of the blue point line, we can certify that $\rho$ is entangled by measuring $\langle\psi_1|\rho|\psi_1\rangle$ and $\langle\psi_2|\rho|\psi_2\rangle$ at the same time.

As a sharp comparison, however, if we choose $|\psi_1\rangle$ and $|\psi_2\rangle$ to be two orthogonal maximally entangled states as in FIG. \ref{fig:figure_good.pdf}(b), we will find that the red line and the blue point line coincide completely, i.e., $f_{|\psi_1\rangle,|\psi_2\rangle}(\text{SEP}) = f_{|\psi_1\rangle,|\psi_2\rangle}(U_2) = [1, 1/d]^2$. This clearly shows that the new approach offers no advantage over original fidelity-based entanglement witnesses. 

\section{An algorithm for picking up optimal $\ket{\psi_1}$ and $\ket{\psi_2}$}

We have known that using maximally entangled pure states to design 2-tuple fidelity-based entanglement witnesses might not be a good idea. Naturally, a new question is raised: Given a quantum state $\rho$, how should we choose proper $|\psi_1\rangle$ and $|\psi_2\rangle$ to design the optimal 2-tuple fidelity-based entanglement witnesses such that the entanglement of $\tilde\rho(p)$ can be detected for the largest noise strength $p$? In the current section, we will mainly focus on the depolarizing noise. In other words, we would like to solve the following optimization problem:
\begin{equation}\label{OptP2}
\begin{array}{ll}
& \max_{|\psi_1\rangle,|\psi_2\rangle} \min_p  p \\
\text { subject to } & \tilde\rho(p)  \in \tilde W_2(|\psi_1\rangle, |\psi_2\rangle),
\end{array}
\tag{P2}
\end{equation}
where $\tilde\rho(p) = p\frac{I_{d^2}}{d^2}+(1-p)\rho$ and $\tilde W_2(|\psi_1\rangle, |\psi_2\rangle)$ is defined in Eq. \eqref{eq:w2PPT}.

Note that if replacing $\tilde W_2(|\psi_1\rangle, |\psi_2\rangle)$ in  Problem (P2) with $W_2(|\psi_1\rangle, |\psi_2\rangle)$ defined in Eq. \eqref{eq:w2SEP} and solving the problem successfully, we can obtain better $|\psi_1\rangle, |\psi_2\rangle$ to detect the entanglement of $\tilde\rho(p)$ with larger $p$. However, we keep the current formulation of (P2) since it is difficult to numerically characterize $W_2(|\psi_1\rangle, |\psi_2\rangle)$. In fact, since $W_2(|\psi_1\rangle, |\psi_2\rangle) \subseteq \tilde W_2(|\psi_1\rangle, |\psi_2\rangle)$, the optimal $|\psi_1\rangle, |\psi_2\rangle$ obtained by solving Problem (P2) can still be used to detect the entanglement of $\tilde \rho(p)$ when $p$ is smaller than the solution to Problem (P2). Therefore, solving Problem (P2) is still valuable for us.  
Moreover, we would like to point out that Problem (P2) can be naturally extended from the 2-tuple case to the $k$-tuple case, i.e., one can utilize the fidelities between the target state and $k$ different pure entangled states to detect entanglement for more quantum states.

The optimization problem (P2) is a max-min optimization problem, and the inner subproblem is
\begin{equation}\label{OptP3prime}
\begin{array}{ll}
& \min p \\
\text { subject to } & \tilde\rho(p) \in \tilde W_2(|\psi_1\rangle, |\psi_2\rangle),\\
& p \in [0,1].
\end{array}
\tag{P3}
\end{equation}
We denote the solution to (P3) by $\mathcal{F}(\psi_1,\psi_2)$. Then the outer subproblem of (P2) now becomes
\begin{equation}\label{OptP4}
  \mathcal{P}(\rho) = \max_{|\psi_1\rangle,|\psi_2\rangle}\mathcal{F}(\psi_1,\psi_2). \tag{P4}
\end{equation}
Problem (P3) is actually a disciplined parametrized programming problem (DPP), which is a grammar for generating parametrized disciplined convex programs from a set of functions or atoms with known curvature and per-argument monotonicities~\cite{diamond2016cvxpy, agrawal2018rewriting}. For optimization problems satisfying the DPP conditions, Ref.\cite{cvxpylayers2019} provides a method for numerically computing the gradient of the optimal solution with respect to the parameters $\ket{\psi_1},\ket{\psi_2}$, which allows us to calculate the gradient with Cvxpylayer~\cite{cvxpylayers2019,agrawal2020differentiating}.
After that, we can use gradient-based methods to solve Problem (P4). However, if randomly selecting the initial values for $|\psi_1\rangle$ and $|\psi_2\rangle$, we find that the gradient $\frac{\partial \mathcal{F}(\psi_1,\psi_2)}{\partial\psi_i}$ is almost always close to a zero vector. 
To overcome this difficulty, we adjust Problem (P4) as below.

We define a function $\Phi(x;M,|\phi\rangle): \mathbb{C}^{d^2} \rightarrow \mathbb{C}^{d^2\times d^2}$ as
$$
\Phi(x;M,|\phi\rangle) = \left(\frac {x/\|x\| + M|\phi\rangle}{\left\|x/\|x\| + M|\phi\rangle\right\|}\right)\left(\frac {x/\|x\| + M|\phi\rangle}{\left\|x/\|x\| + M|\phi\rangle\right\|}\right)^\dagger,
$$
where $x\in \mathbb{C}^{d^2}$ is the variable, and the state $|\phi\rangle$ and the positive scalar $M$ are the parameters. It can be seen that the larger $M$ is, the closer the output of the function $\Phi$ is to $|\phi\rangle\langle\phi|$. We consider the following problem:
\begin{equation}\label{OptP5}
\begin{aligned}
 & \mathcal{P}(\rho;M,|\phi_1\rangle,|\phi_2\rangle) \\
  = &\max_{x_1,x_2\in \mathbb{C}^{d^2}}\mathcal{F}(\Phi(x_1;M,|\phi_1\rangle),\Phi(x_2;M,|\phi_2\rangle)).
  \end{aligned}
  \tag{P5}
\end{equation}
Recall that our goal is to find $|\psi_1\rangle$ and $|\psi_2\rangle$ such that $\mathcal{F}(\psi_1, \psi_2)$ is maximized.

Note that when $M = 0$, the solution to (P5) is equal to that to (P4); when $M > 0$, the solution to (P5) is smaller or equal than that to (P4). Therefore, (P5) provides us a lower bound for the solution to (P4). Furthermore, an advantage of (P5) is that, when the parameters $M, |\phi_1\rangle, |\phi_2\rangle$ are chosen appropriately, the gradient will not always be $0$, making it possible to solve Problem (P5) by continuously updating the parameters $M, |\phi_1\rangle, |\phi_2\rangle$ with gradient-based optimization methods. Next we can adjust the parameters further to make the solution to Problem (P5) approach that to Problem (P4).

However, note that both of Problems (P4) and (P5) are non-convex, therefore a tyipical gradient-based method to solve them will probably converge to local maxima. To address this problem, we utilize an optimizaiton approach called Variational Generative Optimization Network (VGON) proposed recently by Ref.\cite{zhang2024variational} to find the optimal choices for $|\psi_1\rangle$ and $|\psi_2\rangle$.

Basically, similar to the variational autoencoder, VGON is a variational optimization algorithm based on deep generative networks, and has demonstrated a wide applicability and a good efficiency in various quantum tasks~\cite{zhang2024variational}.
Specifically, VGON consists of a multi-layer encoder network, a multi-layer decoder network, and a latent space connecting the encoder and decoder parts. The network processes initial variables through variational optimization to find the optimal solution. When dealing with high-dimensional and complex quantum problems, VGON can avoid getting trapped in local minima, thus can fulfill global optimizations very effectively. More details on VGON can be seen in Ref.~\cite{zhang2024variational}.

\begin{figure*}[!ht]
    \centering
    \includegraphics[width=0.9\textwidth]{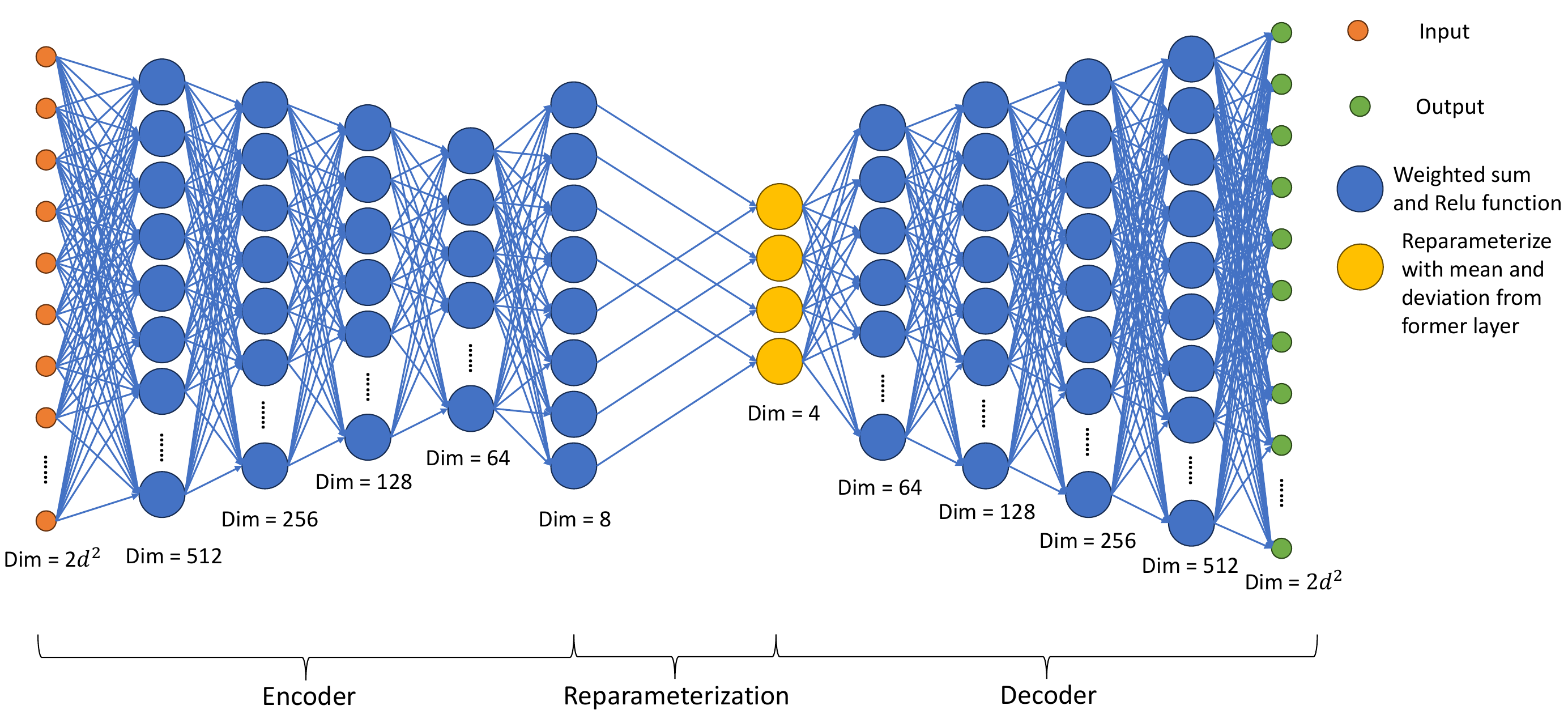}
    \caption{The framework of the VGON network we utilize. Here all the involved quantum states has a local dimension $d = 4$.}
    \label{fig:VAEfigure.png}
\end{figure*}

We construct a VGON network to solve Problem (P5), which will eventually give the solution to (P2). Due to the introduction of the parametric method $\Phi$, the input of VGON is the decision variable $x \in \mathbb{C}^{2d^2}$ of Problem (P5), and the output is a vector of the same dimension as the input. The parameters are in the layers of the linear mapping parts. The specific configuration of the VGON model is shown in FIG. \ref{fig:VAEfigure.png}. Particularly, the hyperparameters in the model are set to be $d = 4$ and $k = 2$, where $d$ represents the dimension of the subsystem of the two-body quantum state $\rho$, and $k$ represents the number of fidelities in designing entanglement witnesses.

We define the computational process of the VGON network depicted above as a ``function" $(\hat{x}_1, \hat{x}_2) = \mathcal{V}(x_1, x_2)$, where $x_1, x_2 \in \mathbb{C}^{d^2}$.
It should be emphasized that due to the stochasticity involved in the reparameterization process, the neural network is not a strict ``function". For the convenience of later discussions, we let $\hat{x}_i = \mathcal{V}_i(x_1, x_2), i = 1,2$.

The definition of our loss function is as follows:
$$
\begin{aligned}
& f_{\mathrm{loss}}(x_1,x_2) \\
= & -\mathcal{F}_2(\Phi(\mathcal{V}_1(x_1);M,|\phi_1\rangle),\Phi(\mathcal{V}_2(x_2);M,|\phi_2\rangle))\\
& + 0.001KL_{div},
\end{aligned}
$$
where $KL_{\text{div}}$ is the Kullback-Leibler divergence between the distribution $N(\text{mean}, \text{var}^2)$ generated in the reparameterization process in neural networks and the standard normal distribution.

Given the initial values of $M, |\phi_1\rangle, |\phi_2\rangle$, the working process of the VGON network is as follows:

\begin{enumerate}
  \item Random generate $((x_1^{(j)},x_2^{(j)})^T$, where $j = 1,2,\cdots,N$. Divide $N$ samples into several groups based on the chosen batch size.
  \item For each iteration, a batch of size $|\Gamma|$ is taken with sample vectors $(x_1^{(j)},x_2^{(j)})^T$, where $j \in \Gamma$. Update the parameters in the VGON network using the gradient information of the objective function $\frac{1}{|\Gamma|}\sum_{j\in \Gamma}f_{\mathrm{loss}}(x_1^{(j)},x_2^{(j)})$.
  \item For each iteration, take a batch of size $|\Gamma|$ with sample vectors $(x_1^{(j)},x_2^{(j)})^T$, where $j \in \Gamma$. Update the parameters in the VGON network using the gradient information of the objective function $\frac{1}{|\Gamma|}\sum_{j\in \Gamma}f_{\mathrm{loss}}(x_1^{(j)},x_2^{(j)})$.
  \item Update the values of $|\phi_1\rangle, |\phi_2\rangle$ by the output of the VGON network. The new value of $M$ is updated by reducing it at a specific rate. Then we can construct a new VGON and return to Step 1.

\end{enumerate}

In the rest of the current section, we will numerically demonstrate that, on many specific examples of target quantum states $\rho$, VGON can effectively find $|\psi_1\rangle, |\psi_2\rangle$ such that the corresponding 2-tuple fidelity-based entanglement witnesses can still detect the entanglement even when high level noise transfers $\rho$ to unfaithfully entangled states.

\subsection{The case of rank-2 quantum states}

For the first example, we will reexamine the quantum state $\rho$ discussed in Section IV, which is
\begin{equation*}
\rho = q_1 |\phi_1\rangle\langle\phi_1| + q_2 |\phi_2\rangle\langle\phi_2|,
\end{equation*}
where $|\phi_1\rangle = \frac{1}{\sqrt{d}}\sum_{i=1}^d |ii\rangle$, $|\phi_2\rangle = |12\rangle$, $q_1+q_2=1$, and $q_1\in\{0.1, 0.2, \cdots, 1.0\}$. Again, here the noise model we choose is the depolarizing noise, which means the quantum state we study is actually $\tilde{\rho}(p) = p\frac{I_{d^2}}{d^2} + (1-p)\rho$.
However, instead of directly letting $|\psi_i\rangle = |\phi_i\rangle, i = 1,2$, we now use VGON to pick up $|\psi_i\rangle$ to construct better 2-tuple fidelity-based entanglement witnesses.

Recall that for the case $|\psi_i\rangle = |\phi_i\rangle, i = 1,2$, by solving (P3) the value of $\mathcal{F}(\phi_1,\phi_2)$ can be obtained as in TABLE \ref{table:result_diffdim1}, where it can be seen that if $p < \mathcal{F}(\phi_1,\phi_2)$, $\tilde\rho(p)$ can be detected to be entangled. As a comparision, the results given by optimizing $|\psi_i\rangle$ with VGON are listed in TABLE \ref{table:CNN1}. It can be seen that the entanglement witness optimized by VGON has better performance.

Let us take the case that $q_1 = 0.1$ as an example. According to TABLE \ref{table:result_diffdim1}, if the noise $p < 0.1221$, the entanglement witness constructed by  letting $|\psi_i\rangle = |\phi_i\rangle$ can detect the entanglement of $\tilde \rho(p)$. Meanwhile, in TABLE \ref{table:CNN1} we can see that the entanglement witness optimized by VGON can achieve this as long as $p < 0.2056$, implying that the latter is more powerful. This example clearly shows that the VGON approach is valuable in constructing high-quality 2-tuple fidelity-based entanglement witnesses. 

\begin{table*}[htb] \scriptsize
    \centering
    \caption {The thresholds of $p$ for the quantum state $\tilde{\rho}(p)$ to be separable, unfaithfully entangled, and to be detected by a 2-tuple fidelity-based measurement witness, where $\tilde{\rho}(p)$ is produced by a depolarizing noise on $\rho$. Specifically, $\tilde{\rho}(p)$ is entangled when $p < p_1(\rho)$; $\tilde{\rho}(p)$ is unfaithful when $p \geq p_2(\rho)$; and $\tilde{\rho}(p)$ can be detected as an entangled state by 2-tuple fidelity-based entanglement witnesses when $p < \max_{\psi_1, \psi_2}\mathcal{F}(\psi_1, \psi_2)$. $\tilde{\rho}(p)$ can be detected as an entangled state by measure fidelity with $|\phi_1\rangle$ and $|\phi_2\rangle$ when $p<\mathcal{F}(\phi_1,\phi_2)$.}
    \setlength{\tabcolsep}{4.5mm}
   {
    \begin{tabular}{ccccc}
\hline\hline $q_1$ value & $P_{\mathrm{SEP}}^{\mathrm{inf}}(\rho)$ & $P_{U_2}^{\mathrm{sup}}(\rho)$ & $\mathcal{F}\left(\phi_1, \phi_2\right)$ & $\max _{\psi_1, \psi_2} \mathcal{F}\left(\psi_1, \psi_2\right)$ \\
\hline 0.1 & 0.28572053593 & 0.016448152780114 & 0.122097364715763 & 0.2056039273738861 \\
 0.2 & 0.444444444444 & 0.081603629602727 & 0.241708841754075 & 0.342727929353714 \\
 0.3 & 0.545462587687 & 0.230768062135074 & 0.357739285807916 & 0.45368921756744385 \\
 0.4 & 0.615368743836 & 0.444444697389034 & 0.468484999675455 & 0.5345901250839233 \\
 0.5 & 0.666667459537 & 0.571428655601596 & 0.571428838096836 & 0.6016825437545776 \\
 0.6 & 0.705881979335 & 0.651159274429514 & 0.651154772589985 & 0.6575230956077576 \\
 0.7 & 0.736846727443 & 0.705883784104214 & 0.705882239234862 & \\
 0.8 & 0.761915755888 & 0.745760316861791 & 0.745766868630711 & \\
 0.9 & 0.782613394564 & 0.776131388134855 & 0.776119454256353 & \\
\hline\hline
\end{tabular}}
    \label{table:CNN1}
\end{table*}

\subsection{The GHZ State with the depolarizing noise}

In the second example, we consider the $4$-qubit GHZ state $|\phi\rangle = \frac{1}{\sqrt{2}}(|0000\rangle + |1111\rangle)$. Each of Alice and Bob holds two qubits of $\ket{\phi}$. Let $\rho = |\phi\rangle \langle\phi|$. In addition, we suppose that $\rho$ is affected by the depolarizing noise and become $\tilde{\rho}(p)= p\frac{I_{16}}{16}+(1-p)\rho$.
We would like to investigate that if $p$ goes up continuously from $0$, when $\tilde{\rho}(p)$ becomes unfaithful, and when it becomes unentangled. 

As in the previous example, we use VGON to solve these problems, and the results are shown in TABLE \ref{table:CNN2}. It can be seen that, when $p \geq 0.57143$, $\tilde \rho(p)$ is an unfaithful state, which means fidelity-based entanglement witness fail to detect the entanglement. However, if we use 2-tuple fidelity-based entanglement wintesses, the entanglement can be detected when $p < 0.59082$. Furthermore, if we construct 4-tuple fidelity-based entanglement wintesses with the help of VGON, their performance can be further improved, i.e., $\tilde \rho(p)$ can be detected to be entangled when $p < 0.71114$. 

Additionally, we would like to point out that when solving Problem (P4), the VGON algorithm can be replaced by the Broyden–Fletcher–Goldfarb–Shanno (BFGS) algorithm~\cite{broyden1970convergence, fletcher1970new, goldfarb1970family, shanno1970conditioning}, a commonly used quasi-Newton method (therefore requiring only gradient information), or the Adam algorithm~\cite{kingma2014adam}, a gradient-based algorithm commonly used in machine learning. If using the BFGS algorithm, the entanglement can be detected when  $p<0.551724$; if using the Adam algorithm, the entanglement can be detected when $p<0.576437$. Recall that the corresponding threshold given by VGON is $0.59082$, implying the advantage of VGON over these two algorithms in designing entanglement witnesses.

\begin{table*}[htb] \scriptsize
    \centering
    \caption {The thresholds of $p$ for the quantum state $\tilde{\rho}(p)$ to be separable, unfaithfully entangled, and to be detected by a $k$-tuple fidelity-based entanglement witness, where $\tilde{\rho}(p)$ is produced by a depolarizing noise on $\rho$. The meaning of $P_{\mathrm{SEP}}^{\mathrm{inf}}, P_{U_2}^{\mathrm{sup}}$ is the same as in TABLE \ref{table:CNN1}. $\tilde{\rho}(p)$ can be detected as entangled by a $k$-tuple fidelity-based entanglement witness when $p < \max_{\psi_1, \cdots,\psi_k}\mathcal{F}(\psi_1, \cdots,\psi_k)$. It can be observed that when the number of fidelities increases, we can certify entanglement for more quantum states.}
    \setlength{\tabcolsep}{4.5mm}
   {
    \begin{tabular}{ccccc}
    \hline
    \hline
$P_{\mathrm{SEP}}^{\mathrm{inf}}(\rho)$ & $P_{U_2}^{\mathrm{sup}}(\rho)$ & $\max _{\psi_1, \psi_2} \mathcal{F}\left(\psi_1, \psi_2\right)$ & $\max _{\psi_1, \cdots, \psi_4} \mathcal{F}\left(\psi_1, \cdots, \psi_4\right)$ & $\max _{\psi_1, \cdots, \psi_8} \mathcal{F}\left(\psi_1, \cdots, \psi_8\right)$ \\
\hline 0.88889 & 0.57143 & 0.59082 & 0.71114594 & 0.80130124\\
\hline
\hline

\end{tabular}}
    \label{table:CNN2}
\end{table*}

\subsection{Heisenberg XY model}

In this subsection, {we will try to detect entanglement for the ground state and the first excited state of the Hamiltonian for the Heisenberg XY model} \cite{wang2001entanglement, lieb1961two}. Here we suppose that these quantum states are affected by both the depolarizing noise and the dephasing noise.

Specifically, the Hamiltonian for the Heisenberg XY model can be written as
\begin{equation}
H = -J \sum_{i=1}^{N} \left( \frac{1+\gamma}{2} \sigma_i^x \sigma_{i+1}^x + \frac{1-\gamma}{2} \sigma_i^y \sigma_{i+1}^y \right) - h \sum_{i=1}^{N} \sigma_i^z.
\end{equation}
The setting we consider here is as follows: The number of qubits is $N=4$, the exchange interaction constant $J=1$, the gamma parameter $\gamma = 0.5$, and the external magnetic field $h=0.5$. 

We denote the ground state and the first excited state of $H$ as $|\phi_1\rangle$ and $|\phi_2\rangle$, respectively. Then we consider the quantum state of form
\begin{equation*}
    \rho = q_1|\phi_1\rangle\langle\phi_1| + q_2|\phi_2\rangle\langle\phi_2|,
\end{equation*}
where $q_1 = 0.7$ and $q_2 = 0.3$. Our task is, under the influence of quantum noise of strength $p$, to find out when the quantum state $\tilde\rho(p)$ is separable, when it can be certified as entangled by fidelity-based entanglement witness (i.e., faithful), and when it can be detected as entangled by our modified entanglement witnesses. Again, we solve the above problems with VGON, and the  results are listed in TABLE \ref{table:CNN3}. 

We can see that regardless of the noise model, our modified entanglement witnesses are always more powerful than orginal fidelity-based entanglement witnesses in detecting entanglement, except for the case that the number of fidelities is 2 and the noise model is the depolarizing noise. Even for the depolarizing noise, when we increase the number of fidelities further, a large proportion of the unfaithful states (among various strengths of noise) can be detected to be entangled.

\begin{table*}[htb] \scriptsize
    \centering
    \caption {The thresholds of $p$ for the quantum state $\tilde{\rho}(p)$ to be separable ($p_1(\rho)$), unfaithfully entangled ($p_2(\rho)$), and to be detected by a $k$-tuple fidelity-based entanglement witness, where $\tilde{\rho}(p)$ is produced by a depolarizing noise or a dephasing noise on $\rho$.}
    \setlength{\tabcolsep}{4.5mm}
   {
    \begin{tabular}{c|ccccc}
\hline
\hline
Noise model &$p_1(\rho)$ & $p_2(\rho)$ & $\max _{\psi_1, \psi_2} \mathcal{F}\left(\psi_1, \psi_2\right)$ & $\max _{\psi_1, \cdots, \psi_4} \mathcal{F}\left(\psi_1, \cdots, \psi_4\right)$ & $\max _{\psi_1, \cdots, \psi_8} \mathcal{F}\left(\psi_1, \cdots, \psi_8\right)$ \\
\hline  depolarizing & 0.726514 & 0.506540 & 0.504318 & 0.533054 & 0.723174\\
dephasing & 0.882348 & 0.363996 & 0.383895 & 0.604390 & 0.863348\\
\hline
\hline
\end{tabular}}
    \label{table:CNN3}
\end{table*}

\subsection{Low-rank random quantum states mixed with the depolarizing noise}

In this subsection, we will randomly generate low-rank quantum states, and then detect their entanglement under the impact of quantum noise with $2$-tuple fidelity-based entanglement witnesses. These examples provide more convincing evidence showing that the modified entanglement witnesses enjoy an apparent advantage over  original fidelity-based entanglement witnesses.

More specifically, we randomly generate $100$ entangled quantum states $\rho$ according to the Haar measure with rank $4$, and all of them are faithfully entangled state. Then we suppose these quantum states are affected by quantum noise with strength $p$. Next with VGON we calculate or bound the following quantities for each quantum state: The minimum value of $p$ such that $\tilde\rho(p)$ is in PPT, i.e., $P_{\mathrm{SEP}}^{\inf}(\rho)$, the minimum value of $p$ such that $\tilde\rho(p)\in \widetilde{U}_2$ (the inner approximation of unfaithful), i.e., $P_{U_2}^{\sup}(\rho)$, and the maximum value of $p$ such that our modified entanglement witnesses can detect $\tilde\rho(p)$ to be entangled, i.e., $\max\mathcal{F}(\psi_1,\psi_2)$.

\begin{figure}[!ht]
    \centering
    \includegraphics[width=0.45\textwidth]{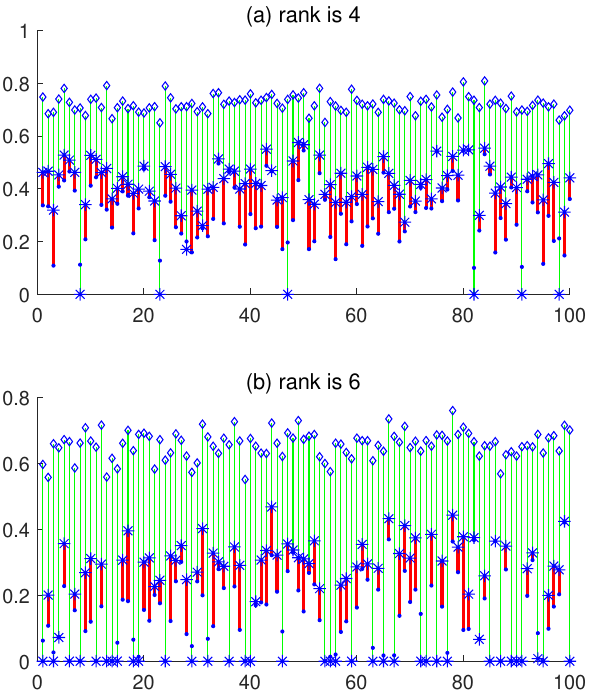}
    \caption{The entanglement of random quantum states. Each point on the X-axis represents a specific quantum state, and in (a) the common rank of the density matrices is $4$, in (b) the common rank is $6$. The Y-axis represents the proportion of mixed-in noise. The symbol ``$\diamond$" in the figure indicates the threshold of $p$ for $P_{\mathrm{SEP}}^{\inf}(\rho)$ (the state is separable when $p$ is larger than this value), the symbol ``$\cdot$" indicates the threshold of $p$ for $P_{U_2}^{\sup}(\rho)$ (the state is unfaithful when $p$ is larger than this value), and the symbol ``$\ast$" indicates the value of $\max\mathcal{F}(\psi_1,\psi_2)$ (the state can be detected as entangled by a 2-tuple fidelity-based entanglement witnesses when $p$ is smaller than this value). From (a) we observe that, among the 100 quantum states generated randomly, 93 ones become unfaithful under the impact of noise but can be identified as entangled by 2-tuple fidelity-based entanglement witnesses. From (b) we observe that, among the 80 faithful quantum states generated randomly, 61 ones become unfaithful under the impact of noise but can be identified as entangled by the modified entanglement witnesses.}
    \label{fig:rank4real.pdf}
\end{figure}

Our numerical calculations show that, on average the modified entanglement witnesses are 0.3433 times more resistant to the depolarizing noise than original fidelity-based entanglement witnesses, i.e., if the threshold of noise strength under which the latter can detect the entanglement of $\tilde\rho(p)$ is $p_0$, the corresponding threshold for the former is roughly $1.3433p_0$ on average.

More details can be seen in FIG. \ref{fig:rank4real.pdf}(a). Here each point on the X-axis represents a specific quantum state. The Y-axis represents the strength of noise. The symbol ``$\diamond$" in the figure indicates the values of $p$ corresponding to the boundary point of PPT, i.e., $P_{\mathrm{SEP}}^{\inf}(\rho)$ (the underlying state is separable when $p$ is larger than this value), the symbol ``$\cdot$" indicates the corresponding value of $p$ for $\widetilde{U}_2$, i.e., $P_{U_2}^{\sup}(\rho)$ (the underlying state is unfaithful when $p$ is larger than this value), and the symbol ``$\ast$" indicates the value of $\max\mathcal{F}(\psi_1,\psi_2)$, which is the optimal solution to Problem (P3) among all possible $\ket{\psi_1}$ and $\ket{\psi_2}$ given by VGON (the underlying state can be detected as entangled by a 2-tuple fidelity-based entanglement witness when $p$ is smaller than this value). 
When $\max\mathcal{F}(\psi_1,\psi_2)>P_{U_2}^{\sup}(\rho)$, if the noise strength $p$ is in the interval $(P_{U_2}^{\sup}(\rho),\max\mathcal{F}(\psi_1,\psi_2))$ (the interval is drawn in red in FIG. \ref{fig:rank4real.pdf}), our modified entanglement witnesses can detect the underlying entanglement but any original fidelity-based entanglement witnesses fail. Among the 100 quantum states we generate, 93 ones can exhibit nontrivial red intervals, implying that the modified entanglement witnesses are more powerful in certifying entanglement.

In addition, we also generate another set of 100 random quantum states with rank $6$ and repeat the above numerical calculations. The results are listed in FIG. \ref{fig:rank4real.pdf}(b), where we can observe that among the 100 quantum states, 20 ones are already unfaithful. Therefore, our discussion will mainly focus on the remaining 80 faithfully entangled quantum states. On these quantum states, we observe that 61 of them can be certified as entangled by our modified entanglement witnesses and cannot by any original fidelity-based entanglement witnesses when the noise strengths $p$ are in the red intervals in FIG. \ref{fig:rank4real.pdf}(b).

Overall, in this case the modified entanglement witnesses are $0.5386$ times more resistant to the depolarizing noise than original fidelity-based entanglement witnesses on average. 

\section{conclusion}

In this paper, we introduce an effective modification for fidelity-based entanglement witnesses, a most popular approach to detect entanglement in modern quantum experiments, such that unfaithful entanglement, a kind of entanglement that cannot be detected by any original fidelity-based entanglement witnesses, can now be detected. For this, we first give a theoretical analysis for the effect of quantum noise on the unfaithfulness of quantum states. Based on these characterizations, we then show that the combination of multiple fidelities can detect unfaithful entanglement caused by quantum noise in many cases. We theoretically analyze the mathematical structure of these new entanglement witnesses, and find that the experience of designing the optimal original fidelity-based entanglement witnesses does not work any more. In addition, using VGON~\cite{zhang2024variational} and Cvxpylayer~\cite{cvxpylayers2019,agrawal2020differentiating}, we have proposed an algorithm that can be utilized to optimize our modified fidelity-based entanglement witnesses, whose effectiveness has been confirmed by quite a few nontrivial examples. Considering the facts that unfaithfully entangled quantum states are common in real-life quantum experiments, and that the modified fidelity-based entanglement witnesses have similar physical implementations with original fidelity-based entanglement witnesses, we believe that the modified entanglement witnesses we propose remove a fundamental drawback of original fidelity-based entanglement witnesses, and can be widely applied experimentally to certify the existence of entanglement. 

\begin{acknowledgements}
We thank Xiaodie Lin for valuable discussions. Ruiqi Zhang and Zhaohui Wei were supported in part by the National Natural Science Foundation of China under Grant 62272259 and Grant 62332009; and in part by Beijing Natural Science Foundation under Grant Z220002.
\end{acknowledgements}

\bibliographystyle{apsrev4-1}
\bibliography{main_detect_entanglement}

\end{document}